\documentclass[pdflatex,sn-mathphys-num]{sn-jnl}% Math and Physical Sciences Numbered Reference Style
\usepackage{graphicx}%
\usepackage{multirow}%
\usepackage{amsmath,amssymb,amsfonts}%
\usepackage{amsthm}%
\usepackage{mathrsfs}%
\usepackage[title]{appendix}%
\usepackage{xcolor}%
\usepackage{textcomp}%
\usepackage{manyfoot}%
\usepackage{booktabs}%
\usepackage{algorithm}%
\usepackage{algorithmicx}%
\usepackage{algpseudocode}%
\usepackage{listings}%
\usepackage[numbers]{natbib}

\theoremstyle{thmstyleone}%
\theoremstyle{thmstyletwo}%

\theoremstyle{thmstylethree}%

\def\ergs{\ifmmode \mathrm{erg\hspace{1mm}s}^{-1} \else erg s$^{-1}$\fi}

\def\micron{\ifmmode \mu\mathrm{m} \else $\mu$m\fi}
\def\msun{\ifmmode \mathrm{M}_{\odot} \else M$_{\odot}$\fi}
\def\msunyr{\ifmmode \mathrm{M}_{\odot} \hspace{1mm}{\rm yr}^{-1} \else $\mathrm{M}_{\odot}$ yr$^{-1}$\fi}
\def\zsun{\ifmmode Z_{\odot} \else Z$_{\odot}$\fi}
\def\lsun{\ifmmode L_{\odot} \else L$_{\odot}$\fi}
\def\mstar{\ifmmode \mathrm{M}_{\star} \else M$_{\star}$\fi}

\begin{document}

\title[Article Title]{High-z galaxies with the JWST and the ELT: Toward Ever-finer Detail}


\author*[1]{\fnm{Eros} \sur{Vanzella}}\email{eros.vanzella@inaf.it}



\affil*[1]{\orgdiv{Osservatorio di Astrofisica e Scienza dello Spazio di Bologna}, \orgname{INAF}, \orgaddress{\street{via Gobetti 93/3}, \city{Bologna}, \postcode{I-40129}, \state{Italy}}}

%\affil[2]{\orgdiv{Department}, \orgname{Organization}, \orgaddress{\street{Street}, \city{City}, \postcode{10587}, \state{State}, \country{Country}}}

%\affil[3]{\orgdiv{Department}, \orgname{Organization}, \orgaddress{\street{Street}, \city{City}, \postcode{610101}, \state{State}, \country{Country}}}


\abstract{
\unboldmath
The exploration of the early Universe is being transformed by the \textit{James Webb Space Telescope} (\textit{JWST}), which delivers unprecedented angular resolution at infrared wavelengths and opens a unique window redward of the \textit{K} band ($2~\mu\mathrm{m}$). Thanks to \textit{NIRCam}, \textit{NIRISS}, \textit{NIRSpec}, and \textit{MIRI} instruments, which provide both imaging and spectroscopy with exquisite efficiency, new classes of sources have emerged within the first years of operations. The first half-Gyr of cosmic time is now routinely probed, revealing massive blue/red galaxies and a population of Active Galactic Nuclei (AGN) appearing as “little red dots” together with a rest-frame near-infrared/optical view of sources across the reionization and post-reionization epochs. Angular resolution will remain pivotal in the 2030s $–$ 2040s, when extreme adaptive optics (AO) facilities will be deployed on both (8 $–$ 10)m (e.g., the VLT Multi-Conjugate-Adaptive-Optics (MCAO) $-$ Assisted Visible Imager and Spectrograph, \textit{MAVIS}) and on extremely large telescopes, like the 39m ELT (e.g., Multi-conjugate adaptive Optics Relay For ELT Observation, \textit{MORFEO}). Operating at the diffraction limit, these facilities will improve JWST’s resolution, with ELT achieving a factor of $\simeq 6$ smaller Point-Spread-Function (PSF). An ELT diffraction-limited PSF (with a Full Width Half Maximum, FWHM~$\approx 8-12$ mas) in the near infrared will resolve spatial scales $<100$ pc at any redshift (z $\lesssim 18$), revealing abundant star-forming clumps with sufficient sensitivity. Leveraging gravitational lensing as a cosmic telescope, even with moderate magnification factors ($\mu \sim 4$–$8$), diffraction-limited 8m and 39m telescopes will probe physical scales $\lesssim 25$ pc, enabling systematic studies of star formation down to star-cluster scale at cosmological distances. Such observations are poised to become routine in the 2030s $-$ 2040s.
}

\maketitle
\section{Introduction}\label{sec1}

The advent of the \textit{James Webb Space Telescope} (\textit{JWST}) is revolutionising our view of the high–redshift Universe. Its exquisite angular resolution ($\sim 60$ mas at 1-2 $\mu m$ and $\sim 100$ mas at 
3 $\mu m$) and its unique access to wavelengths redward of the K band to $\sim 5~\mu m$ with \textit{NIRCam} and to $\sim 27~\mu m$ with \textit{MIRI}, along with the micro-shutter assembly (MSA), slitless and integral–field spectroscopy with \textit{NIRSpec}, \textit{NIRISS} and \textit{NIRCam}, have already yielded a series of discoveries, several of them unexpected. A (non–exhaustive) list includes: 
\noindent (1) efficient photometric selection and rapid spectroscopic confirmation of galaxies and AGN up to $z\simeq 14.4$ (the highest–redshift source known at the time of writing, \citep{Naidu2025}; \citep[see also,][]{robertsborsani2023, Carniani24a, Castellano24, Robertson24, Curtis_Lake2023NatAs,Donnan2026z13});
\noindent (2) the emergence of bright (M$_{\rm UV} \lesssim -20$), blue galaxies at $z > 9$ within the first half–Gyr after the Big Bang \citep[e.g.,][]{Somerville25, Ferrara_2023_monsters, Ferrara2025, Castellano2023_glass, Castellano24, Whitler2025, Napolitano2025, Finkelstein2024, McLeod2024, Donnan2025, Harikane2023, Harikane2024, Perez-Gonzalez2023, Tang2025, Rui2026, Tang2026}, including also bright red galaxies \citep[][]{Rodighiero2026, Ferrara2026}, and similarly luminous objects at lower redshift ($z \simeq 2-6$ \citep{marques-chaves2022, Rui_z6_bright}); discovery of an abundant populations of faint galaxies at $z>9$ \citep[e.g.,][]{Whitler2025_LFs_zGT9, atek2024}; 
\noindent (3) identification of candidate galaxies at $z \sim 15-30$ \citep{Castellano2025_super_highz, Gonzalez2025}, albeit not yet spectroscopically confirmed; 
\noindent (4) the mapping of reionized bubbles during reionization \citep[e.g.,][]{Nikolic2025_reioniz_bubbles, Napolitano2025_bubbles,Umeda2024_bubbles};
\noindent (5) discovery of quenched galaxies up to $z \gtrsim 7$ \citep[e.g.,][]{Santini2025_mature_galaxies, looser2024, de_Graaff2025NatAs}; 
\noindent (6) identification of massive, grand–design spiral/ordered galaxies within $\simeq 1.5$~Gyr of the Big Bang ($z>4$, \citep[e.g.,][]{Xiao2025_spiral,Jones2025_ordered_rotation_highz, Wang2026}); 
\noindent (7) detection with \textit{NIRCam} of HST–dark galaxies at long wavelengths, potentially challenging $\Lambda$~Cold Dark Matter (CDM) interpretations \citep[e.g.,][]{Xiao2024Natur_ultra_massive_galaxies, Kokorev2023_HST_dark};
\noindent (8) discovery of very metal–poor sources that approach pristine star–formation conditions \citep[e.g.,][]{vanzella2023_lap1, Vanzella2024, Fujimoto2025_PopIII, Maiolino2024_PopIII, Maiolino2026, Willott2025, Korber2026, Vanzella2026};
\noindent (9) emergence of a new class of compact sources dubbed ``Little Red Dots'' (LRD), from low redshift ($z \simeq 1$)  to early Universe ($z\sim 10$), showing nuclear activity with elusive host galaxies \citep[e.g.,][]{Furtak2023_LRD, Loiacono2025_LRD}, nearby companions \citep[e.g.,][]{Yanagisawa2026}, high-density/large Balmer break conditions \citep[e.g., ][]{Matthee2026lrd,Naidu2025lrd} and possible link with globular cluster formation \citep[e.g.,][]{Chisholm2026}, see also discussions by \citep[][]{Pablo_LRDs2026, Madau2026lbd_obscured, Madau2026broad_lines} and a review by \citep{inayoshi2025}; 
\noindent (10) galaxies/AGNs with significantly enhanced nitrogen abundances (N–emitters) despite relatively low oxygen abundances \citep[e.g.,][and references therein]{Ji2025, Schaerer2024, Marques_Chaves2024_N, Berg2025_N, Zhu2025_N_enhanced_and_AGN, Morel_schaerer2025_N_enhanced, Schaerer2026}, likely linked to sites of globular cluster formation \citep[e.g.,][]{Schaerer2025_N_and_GCs}; and
\noindent (11) in combination with gravitational lensing, the unveiling of a rich population of massive star clusters across cosmic time, exhibiting properties rarely observed in the local Universe \citep[e.g.,][]{Messa2025, adamo_sparkelr_2023, adamo2024a, vanzella2_sunrise2023, Vanzella2025, claeyssens24, Claeyssens2026}.
\noindent For more complete reviews of \textit{JWST} 
%results \citep[see also,][]{Adamo2025NatAs, Matthee2025review}.
results see also \citep[][]{Adamo2025NatAs, Matthee2025review}.

\section{From HST to JWST: Bound Star Clusters and Pristine Star Formation at Cosmological Distances
}\label{sec2}

High–redshift star clusters were already identified in the pre–\textit{JWST} era in strongly lensed fields with \textit{Hubble} \citep[e.g.,][]{vanzella2017b, vanzella2017a, vanzella2019, vanzella2022}, along with several parsec–scale star complexes \citep[][]{bouwens21, rigby17, johnson17, mestric2022, welch2023}. The advent of \textit{JWST} has enabled the identification of similar parsec–scale stellar clump regions at lower magnification values and higher redshift \citep[][]{Messa25_CG, messa2024b, claeyssens24, vanzella2_sunrise2023, Mowla2024x, Hsiao24, fujimoto2024, adamo2024}, and has even allowed the detection of relatively old star clusters thanks to its extended wavelength coverage with onboard \textit{NIRCam} and \textit{MIRI} imagers \citep[][]{adamo_sparkelr_2023}. In particular, two aspects have improved significantly after the \textit{JWST} launch: (a) the superb angular resolution, coupled with gravitational lensing, allows us to constrain the radii of compact sources hosted in highly distorted distant galaxies down to a few tens of parsecs (for magnification factors $\mu >10$), and to reach a few–parsec scales at very high magnification ($\mu > 30$); and (b) the rest–frame near–infrared ($1.2\mu m$), optical ($0.7\mu m$), and blue ($0.45\mu m$) wavelengths of sources at $z \sim 3, 6, 10$ are now accessible. This is crucial for a more stable inference of stellar masses and ages, especially when spectroscopic information, such as Balmer emission lines, is available. By combining age, stellar mass, and radius, the dynamical age ($\Pi$) can be derived as the ratio between the age and the crossing time ($\Pi = \rm age / T_{CR}$), the latter inferred from the star cluster’s mass and size \citep[][]{adamo20, gieles11}. The crossing time is the characteristic time it takes a typical star to travel across the cluster and adopting a virial scaling ($\sigma^2 \sim GM/R$) is expressed in megayears as $\rm T_{CR} = 10 \times (\rm R_{eff}^{3} / GM)^{\frac{1}{2}}$, where M and $\rm R_{eff}$ are the star cluster’s mass and effective radius expressed in solar masses and parsec, respectively, and
$\rm G\simeq 0.0045$~pc$^{-3}$~\msun$^{-2}$~Myr$^{-2}$ is the gravitational constant.
Compact sources (e.g., with radius $<10$~pc) showing stellar mass surface density larger than $\sim 1000$ \msun~pc$^{-2}$ and $\Pi > 1$ can be considered bound star clusters: the system has survived multiple crossings, which generally implies it is close to dynamical equilibrium. An object of $10^6$ \msun\ and R$_{\rm eff}=10$ pc will have $\Pi > 1$ if older than 5 Myr. Such objects are now accessible with \textit{JWST} coupled with strong lensing \citep[][]{Messa25_CG, adamo2024}.

The statistical study of high–redshift lensed clumps, including bound star clusters, across $1<z<12$ is still in its infancy \citep[][]{messa2024, Messa2025, adamo2024, Vanzella2025, Messa2025, claeyssens24, Bradac2025_z11, Mowla2024Natur_firefly_sparkle, Nakane2025}. Nevertheless, a clear pathway has already emerged, opening the possibility to infer key quantities such as the star–cluster mass function, characterized by a power–law shape with possible truncations at both the high and low mass ends, and the cluster formation efficiency, CFE, i.e., the fraction of star formation that occurs in bound star clusters over a given time interval (equivalently the ratio of the cluster formation rate to the star formation rate of the galaxy over that interval of time \citep[][]{adamo17}). Until recently, both quantities were largely limited to studies in the local Universe. The shape of the mass function and its truncation masses have been found to likely depend on environmental properties such as the star–formation surface density of the host galaxy, and they will likely evolve with cosmic time, a dependence which remains to be observationally explored. Likewise, the mode of star formation may positively correlate with a high CFE: as the Universe becomes denser at earlier times, conditions may favor prominent star–cluster formation with a mass–function slope and truncation masses that are currently unknown. CFE has already been inferred for a few lensed high-z galaxies, the Sunburst arc \citep[][]{vanzella2022}, the Sunrise arc \citep[][]{vanzella2_sunrise2023} and the Cosmic Gems galaxy \citep[][]{Vanzella2025, adamo2024a}, yielding notably high values, $>50-80\%$.

Studying the star–cluster mass function bears directly on the nature of star formation across cosmic time and is tightly linked to (a) globular–cluster formation, as recent simulations show \citep[e.g.,][]{calura2025, pascale2025, Nebrin2025, Andersson2024_preSN, krumholz2019, Garcia25, Sugimura2024} (\citep[see also,][]{renzini17, pozzetti19}), as well as to (b) cosmic hydrogen reionization \citep[e.g.,][]{he20_ricotti, Menon2025_LyC_star_clusters, katz13, schaerer11}. The connection between bound star clusters at early times and local ancient globular clusters will be eventually established with simulations. It is worth noting that a parsec–scale, young bound cluster identified at $z\sim 6-10$ would, if left unperturbed, likely survive to the present day and resemble a $\simeq 13$ Gyr–old star cluster, likely inheriting the characteristic properties of local globular or nuclear star clusters (size, density, metallicity). It is worth recalling the connection between dense star clusters and intermediate-mass black holes (IMBHs). One viable pathway for the formation and early growth of supermassive black holes is the assembly of IMBHs. High redshift nuclear star clusters provide ideal nurseries for this process: their very high stellar densities substantially enhance the rate of stellar–mass black hole mergers and/or stellar collisions in their interiors, paving the way for IMBH seeds of $10^3 - 10^6$~\msun\ \citep[][]{Rantala2025,Rantala2026, Garcia2025OJAp_Nuclear_star_clusters, Kritos2025ApJ_NSC_BH, adamo2024, Katz2015MNRAS_IMBH_NSC,Schneider2023MNRAS_surprise}, see also \citep[][]{Neumayer2020A&ARv_NSC} for a review.

\vspace{2mm}

\noindent {\bf Hydrogen reionization}. The time and spatial evolution of the volume filling fraction of ionized hydrogen $\rm Q_{II}(z)$ is the key quantity in reionization models \citep[][]{Dayal2018PhR}. It can be written as a balance between ionizations and recombinations: 
\[
\frac{\mathrm{d}Q_{\mathrm{H\,II}}}{\mathrm{d}t}
=
\frac{\dot{n}_{\mathrm{ion}}}
     {\langle n_{\mathrm{H}}\rangle}
-
\frac{Q_{\mathrm{H\,II}}}
     {t_{\mathrm{rec}}}.
\]

%$\frac{\rm dQ{II}}{\rm dt} = \frac{\rm \dot n_{ion}}{\rm n_{H}} - \frac{\rm Q{II}}{\rm t_{rec}}$, 
\noindent where $\dot{n}_{\mathrm{ion}}$ is the comoving production-rate density of ionizing photons escaping into the intergalactic medium (IGM), $\langle n_{\mathrm{H}} \rangle$ is the mean comoving hydrogen number density, and $t_{\mathrm{rec}}$ is the effective recombination time
%where the first term represents the production rate of ionizing photons per hydrogen atom in the intergalactic medium (IGM), and the second term accounts for recombinations over the typical timescale $t_{rec}$ 
(which depends on the IGM clumping factor \citep[e.g.,][]{Shull2012ApJ_clumping}). 
%The ionizing–photon production rate can be expressed as $\rm \dot n_{\rm ion} = f_{esc}~\xi_{ion}~\rho_{SFR}$ 
The ionizing-photon production-rate density can be expressed as $\dot{n}_{\mathrm{ion}} = f_{\mathrm{esc}}\,\xi_{\mathrm{ion}}\,\rho_{\mathrm{SFR}}$
\citep[][]{Kuhlen2012_EoR, Robertson2010Nature_EoR}, where the first quantity represents the fraction of escaping ionizing photons into the intergalactic medium, the second the ionizing photon production efficiency (defined as the production rate of hydrogen-ionizing Lyman-continuum photons per unit of intrinsic non-ionizing ultraviolet continuum luminosity) and the third the star formation rate density of the Universe at the given redshift. In a scenario in which reionization is driven by star formation, star clusters naturally enter the problem through $\dot{n}_{\mathrm{ion}}$. In particular, star clusters host massive O–type stars that dominate the stellar ionizing–photon budget \citep[][]{O_stars_in_clusters2012,crowther2016,massive_stars_in_YMC_2017}. Very massive stars (VMSs; $M >100$ \msun) are increasingly recognized as a key ingredient in stellar–evolution and atmosphere models used to reproduce and predict the ultraviolet properties and ionizing–photon production efficiency ($\rm \xi_{ion}$) of star–forming galaxies \citep[e.g.,][]{schaerer2024_VMS}. In addition, the combined (early and/or late) feedback from spatially clustered or sequential (even coeval) cluster formation episodes acts on the interstellar medium, clearing gas and carving ionized channels that promote Lyman–continuum escape ($\lambda < 912$ \AA) and thus the ionization of the intergalactic medium ($\rm f_{esc}$). Hence, star clusters not only produce ionizing photons efficiently, but may also facilitate their escape. The role of star clusters in high–redshift galaxies is therefore pivotal in the reionization process \citep{adamo2024}; \citep[see also,][]{katz13, he20_ricotti, renzini17, Mowla2024Natur_firefly_sparkle, bik18}.

\begin{figure}[h]
\centering
\includegraphics[width=0.99\textwidth]{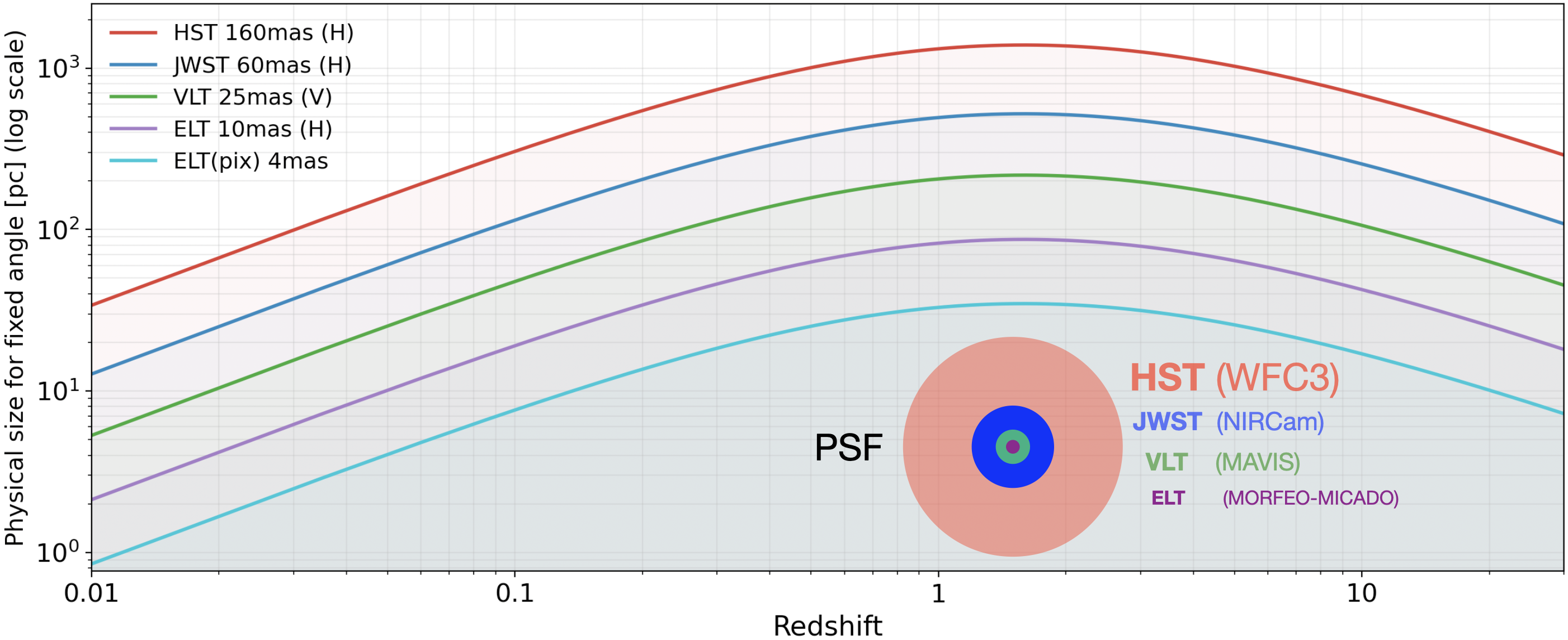}
\caption{Spatial scales probed for different angular resolutions as a function of redshift (color–coded curves); the main facilities discussed in the text are reported in the top-left legend.
\textit{ELT/MORFEO–MICADO} will probe $<100$ pc at any redshift without gravitational lensing (and $<40$ pc per pixel in $\rm MCAO$ mode). The circles at the bottom center schematically indicate the different PSF sizes, with diameters rescaled to the same scale ranging from 160 mas (\textit{HST}) to 10 mas (\textit{ELT}) in the H band; the \textit{VLT/MAVIS} (green) circle is shown for the V band.}\label{fig1}
\end{figure}

\vspace{2mm}

\noindent {\bf Pristine star formation}.
With the advent of \textit{JWST}, the search for metal–free, Population~III (Pop~III) sources has entered a golden era. Although no Pop~III stars have yet been observed directly, their existence is strongly supported by cosmological simulations \citep[][and references therein]{abel2002, Bromm2002, hirano2014, Park2021, klessen2023} and by the discovery of extremely metal–poor halo stars that likely bear the nucleosynthetic imprint of Pop~III enrichment \citep[e.g.,][]{salvadori2007, hartwig2018, vanni2023}. The identification of pristine stars is extremely challenging \citep[][]{katz2022} and currently requires gravitational–lensing amplification. Lensing provides a powerful route to probe pristine stars \citep[][]{rydberg2013, zack2015, vikaeus2022, vanz_popiii} by enhancing (i) spatial contrast and (ii) the signal–to–noise ratio of diagnostic features $-$ albeit at the cost of a reduced effective survey volume at high magnification. If extremely metal–poor (EMP) or metal–free conditions are confined to very small regions (e.g., before pollution by previous or nearby star formation), achieving high spatial contrast becomes essential. Indeed, the expectation that an isolated star complex or cluster of EMP/metal–free stars forms in relative isolation imposes demanding observational requirements \citep[][]{katz2022}.
Despite these difficulties, a number of very metal–poor candidate star–forming complexes have recently been reported at 
$z \simeq 3–8.3$ \citep[][]{Cai2025, vanzella2023_lap1, Vanzella2025_LAP2, Fujimoto2025_PopIII, Morishita2025_AMORE6, Cullen2025_lowZ, Trussler2026}. Access to the rest–frame optical continuum and emission lines enables the use of indirect gas–phase metallicity indicators (e.g., the R3 index; \citealp{Sanders24, nakajima2022}). However, calibrations of such strong–line diagnostics are not yet robust at the lowest metallicities, and these systems may represent the most metal–poor objects currently known in the Universe. Deep spectroscopy will be crucial for decisive progress, particularly to investigate high–ionization, high-density \citep[e.g.,][]{Hsiao2026} and nebular extinction on metal–deficient sources \citep[][]{Trussler2026, Caputi2026pseudo}. He–strong (HeII~1640) emission in faint and possibly pristine metal-deficient objects \citep[e.g., close to GNz11,][]{Maiolino2026,Hubler2026heII, Rusta2026} also represents cases with great potential for ground-based big telescopes \citep[e.g.,][]{POPIII_Nakajima2022, Gonzalez2025_HeII_WR}. See \citep[][]{Venditti2026} for a recent review on the status and perspective on the search for Population III stars.

\begin{figure}[h]
\centering
\includegraphics[width=0.99\textwidth]{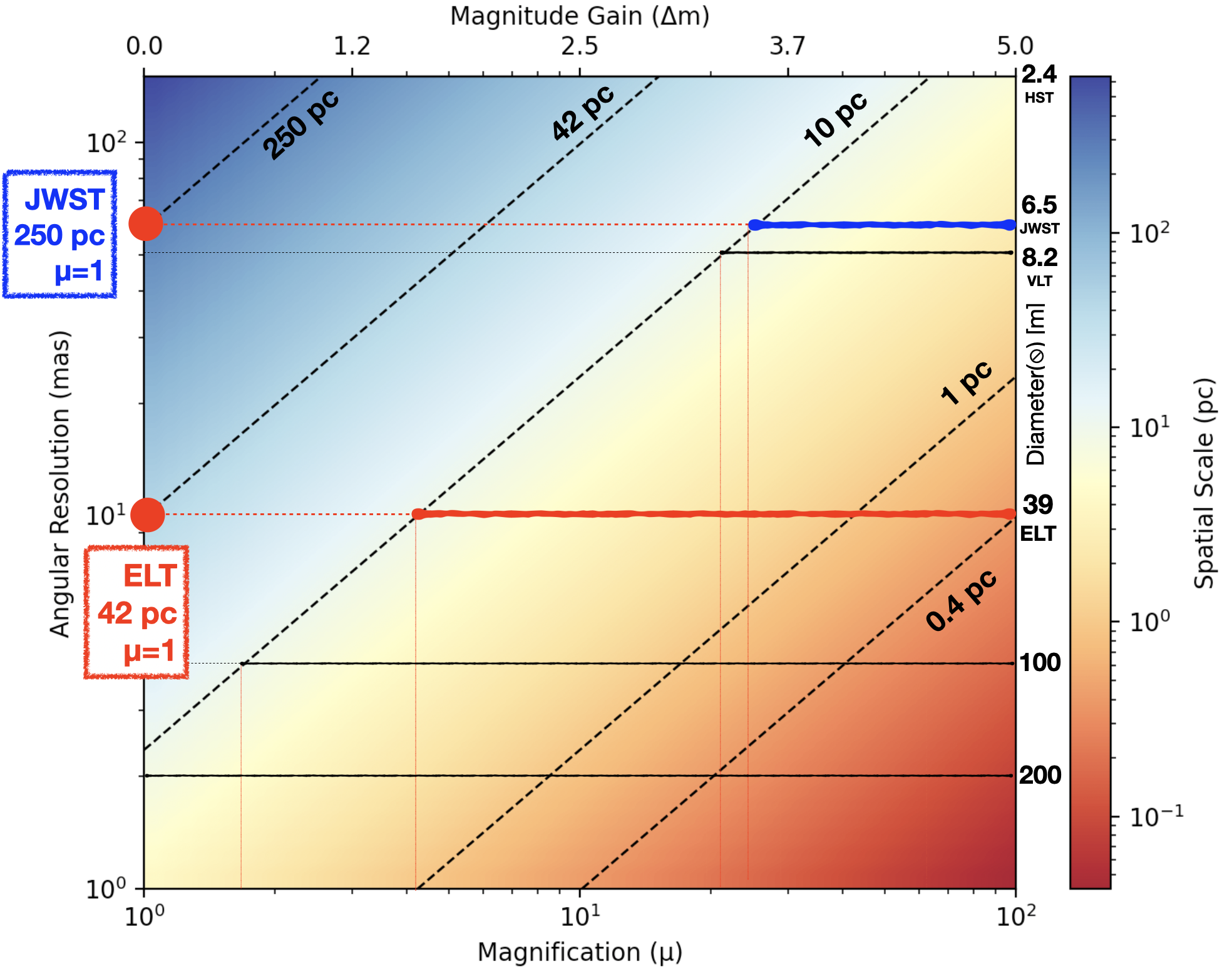}
\caption{Spatial scale (color coded area) as a function of the diffraction limited angular resolution at the given telescope diameter (right Y-axis), and magnification $\mu$, for a compact source placed at redshift 10. The adopted $\mu$ is here generally assumed as tangential(T) or radial(R) component, which dominates the total one, such that $\mu_R \times \mu_T = \mu_{\rm total}$ and $\mu_R \simeq \mu_{\rm total}$ or $\mu_T \simeq \mu_{\rm total}$ (depending on the geometry of the lens). The Y-axis on the left at $\mu=1$ corresponds to no lensing amplification adopting redshift 10: 250(42) parsec are probed with the adopted \textit{JWST}/(\textit{ELT}) FWHM PSF of 60(10) milli-arcsec (in the H-band). Top axis roughly reports the magnitude gain at the corresponding magnification $\mu$ (valid for compact objects). With modest magnification ($3 < \mu < 10$), \textit{ELT} will probe the star cluster regime well below 25 parsec, while at $\mu>30$ sub-parsec scales will be accessed and will represent new science (see the thick red horizontal line).}\label{fig2}
\end{figure}

\section{The Extremely Large Telescope Revolution and the synergy with JWST: Unveiling Bound Star Cluster Populations Across Cosmic Time}\label{sec3}

The \textit{European Extremely Large Telescope} (\textit{ELT}), with its 39m primary mirror diameter, is the facility which will enable a quantum leap in studies of star clusters and pristine star formation in the early Universe. In particular, the Multi-AO Imaging Camera for Deep Observations imager, \textit{MICADO}
%\textit{\href{https://elt.eso.org/instrument/MICADO/}{MICADO}} 
\citep[PI, Rick Davies,][]{MICADO_2024_status}, coupled with the Multi-conjugate adaptive Optics Relay For ELT Observation, \textit{MORFEO}
%\textit{\href{https://elt.eso.org/instrument/MORFEO/}{MORFEO}} 
\citep[PI, Paolo Ciliegi,][]{MORFEO_2024}, will deliver diffraction-limited imaging with $8-12$ milliarcseconds (mas) PSF in the near-infrared J to K band, with Strehl ratios up to 65\% in the K-band, depending on the asterism, the availability of natural guide stars and adopted field of view (FoV). \textit{MORFEO-MICADO} will deliver relatively large AO-corrected FoVs  over tens of arcseconds on a side, from $20'' \times 20''$ up to $51'' \times 51''$ (with 1.5 and/or 4 mas per pixel). 
The AO-corrected \textit{ELT} PSF ($<12$~mas) corresponds to less than 100~pc at any redshift accessible from the ground (see Figure~\ref{fig1}), even 
without gravitational lensing. The $6\times$ improvement relative to the \textit{JWST} H-band resolution 
will reduce the measurable radii of high-redshift sources below 100~pc, and will readily access the star-cluster regime in lensed fields. In fact, cluster–scale sizes ($<25$~pc) will be reachable with modest gravitational lensing magnifications, $\mu=5-8$ (geometry-dependent: assuming here the total magnification, which is the product of the tangential and radial components, is dominated by the highly-magnified component of the two).
%the total magnification is similar to the tangential/radial component). 
Figure~\ref{fig2} summarizes what can be achieved at 
$z=10$ for different telescope diameters and lensing magnifications. As $\mu$ increases, progressively smaller intrinsic scales are probed, effectively mimicking the PSFs of even larger telescopes, up to the equivalent of 100m diameters (already, \textit{JWST} combined with $\mu \simeq 8-10$ delivers \textit{ELT}-like effective resolution). \textit{ELT} observations of moderately lensed galaxies ($\mu \simeq 5-10$) will thus preview the capabilities of future $\gtrsim 100$m-class facilities, while highly magnified targets ($\mu>30$) will push into sub-parsec scales for the first time. Lensing benefits not only angular resolution but also depth: 
for compact sources the magnitude gain is $\Delta m$ = 2.5 log$_{10}(\mu)$. 
A compact source with observed magnitude $H~=~29.5-30$ AB ($5\sigma$ detection achievable in $\simeq 5$~h with \textit{MORFEO–MICADO}) corresponds to an intrinsic 
$H = 32-32.5$ if magnified by $\mu = 10$. Notably, for compact sources without lensing, the exposure time required to reach the same signal-to-noise ratio as a source amplified by a factor $\mu$ scales as $\mu^2$ \citep[e.g.,][]{vanzella2021}: 5h on a lensed compact object with $\mu=10$ delivers roughly the same S/N as 
$\sim 500$ hours without lensing. This, of course, depends on uncertainties in the lens models, which typically grow with magnification \citep[][]{Meneghetti_2017}. 
%However, the reliability of magnification maps $-$ and their error estimates $-$ has improved markedly thanks to the large number of spectroscopic constraints from hundreds of background multiple images at $1<z<10$ and properties of the cluster members \citep[][]{bergamini2023, bergamini_2022_J0416, richard2021, Rihtar2025_lens_model}. 
However, the reliability of magnification maps (and of their associated uncertainty estimates) has improved markedly thanks to the large number of spectroscopic constraints provided by hundreds of multiple images of background sources at $1<z<10$, together with the detailed characterization of cluster members \citep[][]{bergamini2023, bergamini_2022_J0416, richard2021, Rihtar2025_lens_model}. Spectroscopic confirmation is crucial because it securely establishes which images originate from the same background galaxy and accurately determines the source redshift. The positions and redshifts of these multiple images provide strong constraints on the cluster mass distribution, reducing degeneracies in the lens model and improving the determination of the critical curves and magnification factors.
In particular, \textit{VLT/MUSE} spectroscopy and \textit{JWST} imaging/spectroscopy have greatly accelerated this progress, and \textit{ELT} imaging will further 
increase the number of identifiable multiple$-$image clumps, providing an additional boost to lens modeling, possibly delivering thousands of multiple images/clumps (this will require dedicated tools and techniques for properly model the lenses, \citep[e.g.,][]{Lombardi2024_lens}). As a result, in the low$-$magnification regime ($\mu 
\lesssim 10$), statistical errors can be reduced to only a few percent, while at large magnifications, close to critical lines, systematic errors dominates.

Once operational, a natural \textit{ELT} extragalactic science case will be the characterization of star–cluster mass functions in high-z galaxies and the measurement of the CFE across cosmic time \citep[e.g.,][]{messa2024, adamo2024}. This will be a key topic of the 2030s $-$ 2040s. Volume–limited studies in lensed fields will complement unlensed surveys: the former will resolve the internal properties of clumps down to the cluster scale, while the latter will provide statistically robust demographics of clump formation \citep[e.g.,][]{zanella19, ono2024, Sok2025_clumps_CANUCS, Mercier2025_clumps_COSMOSWEBB, de_la_Vega_2025_clumps}.

%\subsection{The JWST-ELT Synergy: Star Cluster Populations Across Cosmic Time}\label{subsec2}

As discussed above, the dynamical age inferred for high-redshift sources depends critically on their size, which will be tightly constrained with diffraction-limited \textit{VLT} (e.g., \textit{MAVIS})
%(e.g., \href{https://www.eso.org/sci/facilities/develop/instruments/MAVIS.html}{\textit{MAVIS}}) 
or \textit{ELT} (\textit{MORFEO-MICADO}) imaging, down to a few dozens of parsecs in blank fields and a few parsecs in lensed fields. By contrast, ages and stellar masses rely predominantly on \textit{JWST}, especially at progressively higher redshift, where the long wavelengths uniquely sample the rest-frame optical spectra via imaging (\textit{NIRCam}/\textit{MIRI}) and spectroscopy (\textit{NIRSpec}/\textit{NIRISS} and \textit{NIRCam} slitless). Consequently, the natural first ELT targets in the early Universe studies will be those already identified and characterised by \textit{JWST} (or soon to be).
Moreover, given the very fine pixel scale of \textit{MORFEO-MICADO} of 4 mas/pix in multi-conjugate adaptive-optics mode, MCAO, diffuse light will typically be lost or heavily buried in the background noise $-$ an effect highlighted by ongoing simulations (\citep[e.g.,][]{Simioni2026} and Scaloni et al in prep. and Mini et al. in prep.). 
The synergy therefore is clear: \textit{JWST} provides a global view and robust stellar–population constraints for the host galaxy, while the \textit{ELT} resolves the star clusters themselves $-$ whose stellar masses and ages will still benefit from \textit{JWST} rest–optical observations. Together, they deliver a continuous, complementary picture of the star–cluster population in its galactic context. Previous and ongoing \textit{JWST} programs (GO and GTO) in both blank and lensed fields are crucial for identifying ideal targets for future ground–based \textit{VLT} and \textit{ELT} facilities with extreme adaptive optics. 
%In particular, the large \JWST\ program \textit{VENUS} (Vast Exploration for Nascent, Unexplored Sources; 300,h, PI Fujimoto) surveys 60 galaxy clusters with NIRCam to a medium depth of $\simeq 28.5$ ($5\sigma$) per band, and NIRSpec for dedicated follow-up spectroscopy, with a focus on the early Universe (first half Gyr and first stars).

\noindent Finally, while \textit{JWST} will continue to identify extremely low-metallicity candidates, \textit{ELT} spectroscopy will be decisive for testing one of the key ultraviolet diagnostics linked to Population~III stars, He\textsc{ii} $\lambda 1640$ \citep[e.g.,][]{schaerer2002, schaerer2003, POPIII_Nakajima2022,Gonzalez2025_HeII_WR}. \textit{MICADO} long-slit spectroscopy can target this emission in compact, lensed sources, and subsequently \textit{HARMONI}, 
%\textit{\href{https://elt.eso.org/instrument/HARMONI/}{HARMONI}}, 
the integral-field spectrograph on the second \textit{MORFEO} port, will map line emission from the target and its surrounding regions.

%\section*{Declarations}

%\subsection*{Funding}

%Not applicable. 

%\subsection*{Availability of data and material}

%Not applicable.

%\subsection*{Competing interests}

%Not applicable.

%\subsection*{Authors' contributions}

%Not applicable.

%\bibliographystyle{naturemag}
\bibliography{sn-bibliography}
\end{document}